\documentclass{pramana}

\usepackage{graphicx,amsmath,bm}

\begin{document}

\title{Simulation of Proton and Carbon-12 Ion Beam for Tumor/Cancer Treatment}

\author{R. Kanishka\textsuperscript{*}}
\affilOne{Center of Excellence in Computational Physics, Department of Physics, University Institute of Science, Chandigarh University, Mohali, Punjab, India 140413\\}


\twocolumn[{

\maketitle

\corres{kanishka.rawat.phy@gmail.com}


\begin{abstract}
Treating cancer is one of the most challenging task in medical sciences. Only limited types of cancer treatments are available as their study is still ongoing. The earlier therapies like radiotherapy with x-rays, chemotherapy are associated with lot of side-effects. One of the most desirable cancer treatment is using particle beam therapies. These therapies are quite less risky than other types of cancer and tumor treatments. In this paper we present the proton and carbon-12 ion beam simulation that can help in tumor and cancer treatment. We simulated the proton and carbon-12 ion beam in water and soft tissue using geant4 toolkit. The protons are observed to have much better energy deposition in the water and soft tissue than carbon-12 ion and gamma photon beams. 
\end{abstract}



}]

\doinum{:}
\artcitid{\#\#\#\#}
\volnum{}
\year{2024}
\pgrange{1--7}
\setcounter{page}{1}
\lp{7}

\section{Introduction}
\label{intro}

Radiation therapy uses radiation dose to kill cancer cells and shrink tumors \cite{citation1}--\cite{citation2}. It is done for the treatment of cancer, used either alone or after combining with other treatments like surgery, chemotherapy, or immunotherapy. Radiation therapy works by damaging the DNA within cancer cells. This damage prevents the cells from dividing and growing, ultimately causing cell death. But the radiation therapy such as x-rays affects both cancerous and normal cells \cite{citation3}--\cite{citation4}. Hence we require some other beam that can kill only cancer cells and not effect the normal cells or tissues. 

Radiotherapy using proton and carbon-12 ion beam is a promising treatment for tumor/cancer. However, understanding the interactions of protons and carbon-12 ion beam with matter is crucial for optimizing treatment efficacy and minimizing side effects \cite{citation5}--\cite{citation11}. The practical work on particle beam therapy requires lot of refinements in dose calculations. A proton dosimetry is a type of particle beam radiation therapy that use protons rather than x-rays to treat cancer. The treatment planning uses accelerators that accelerates the protons and then the dose is given to the patients to treat the cancer. It is a very precise form of treatment that delivers high doses of radiation directly to the tumor/cancer while minimizing damage to surrounding healthy cells, tissues and organs.

Before putting our studies in the practical form, we need to simulate and understand the physics behind the proton and carbon-12 ion beam interactions with water and tissues to refine the dosimetry so that we can avoid any side-effects of beam therapy on human body. Proton therapy involves the use of protons, which are positively charged particles. Unlike x-rays used in traditional radiation therapy, protons have unique properties that make them deposit most of their energy directly in the tumor with minimal exit dose. This phenomenon is known as the Bragg peak. The carbon-12 ion beam shows a similar behaviour.

The protons and carbon-12 ions interaction with matter is characterized by ionization and excitation, Bremsstrahlung, pair production, nuclear stopping and multiple scattering processes \cite{citation12}--\cite{citation18}. As protons travel through matter, they lose energy gradually due to ionization and excitation. The rate of energy loss increases as the proton slows down and near the end of the range, they deposit the majority of their energy in a narrow region, creating a Bragg peak in the dose distribution. The Bragg peak is a characteristic feature of the energy deposition profile of charged particles, particularly protons and heavy ions like carbon-12 ion, as they traverse through matter \cite{citation19}--\cite{citation25}. The depth of the Bragg peak depends on various factors, including the initial energy of the charged particle, the type of particle, and the composition and density of the material it is passing through. For our studies we have simulated the protons, carbon-12 ion beam in water and soft tissue and studied them using geant4 \cite{citation26} simulation toolkit. The paper is organized as follows: methodology have been discussed in section 2, results has been shown in section 3, summary and conclusions have been discussed in the last section.

\section{Methodology}
\label{method}

\begin{figure*}
\centering\includegraphics[height=.35\textheight]{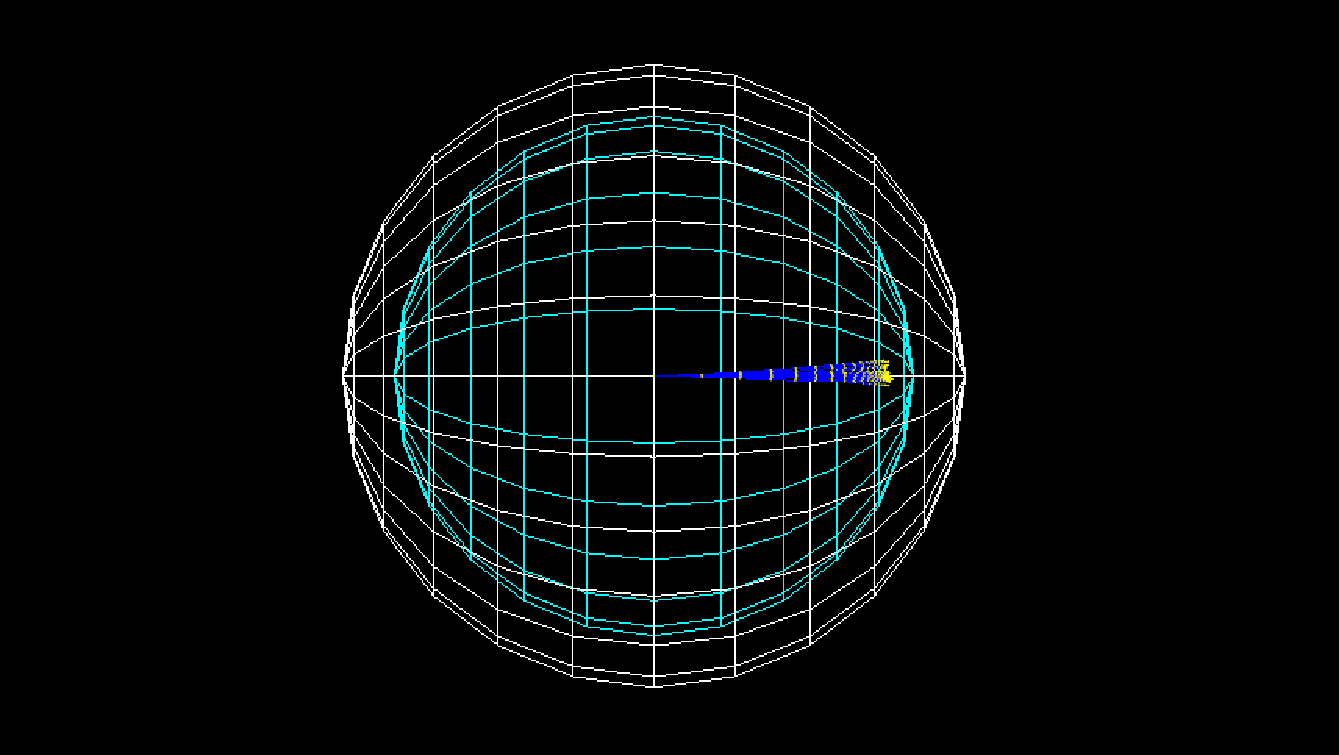}
\caption{The visualization of simulation of interaction of protons in water. Note: The cyan sphere represents water volume. The blue lines show 50 protons shot in water volume, whereas the yellow points shows the ionization process.}\label{figOne}
\end{figure*}

We used geant4 software toolkit \cite{citation26} that is used for the simulation of the passage of particles through matter. It is widely used in high energy physics, space science, medical physics, and radiation protection. The particles when interact with matter, their properties are detected and their path is visualized with geant4 toolkit. The data obtained from geant4 analysis has been further analyzed with ROOT software \cite{citation27} which is a data analysis toolkit used for high energy physics. Since the tumor is a solid mass of tissue mostly and cancer contains mostly water therefore we choose to study the interactions of proton and carbon-12 ion beams with soft tissue/water medium. In the geant4 program we defined the material, its density, volume, physics interactions, type of particles, direction, position and energy. We considered spherical shaped volume filled with water or soft tissue. The particles where they start their journey is called ``prestep'' and where they interact is called ``poststep'' has been defined. We simulate 10,000 particles (protons, carbon-12 ion) of different energies inside the spherical volume of water (4$\pi$ $\times$ $(30)^{3}$ /3) $cm^{3}$. The initial position of protons were set to x = y = z = 0 cm and shot along positive z-direction. The figure~\ref{figOne} shows the geant4 display of simulation of interaction of protons in water medium. We show the simulation of 50 protons of 50 MeV energy that were shot in a water medium. The white sphere represents the world volume which means the outer volume of the tissue/cell and it doesn't participate in any interaction. The main part is the cyan coloured sphere that represents water volume. The protons were kept at their initial position which is x = y = z = 0 cm and shot along the z direction. The physics lists such as PhysListEmStandard, EMLivermore and EMPenelope have been incorporated in our study as these are used for the interactions of protons with matter. The yellow coloured points shows the ionization process. The figure shows that the protons are stopped inside the water after travelling some distance. We show the results in the next section.

\section{Results}
\label{rs}  
For our studies we choose protons and carbon-12 ions interactions with water and soft tissue. As we discussed earlier that protons and carbon-12 ion interact with matter differently than other particles. Here we discuss our results obtained from simulation of protons and carbon-12 ion in water and soft-tissue using geant4 toolkit which have been discussed in the next sub-sections.

\subsection{Simulation of Protons and Carbon-12 Ion Beam}
\label{el}
\begin{figure*}
\centering\includegraphics[height=.35\textheight]{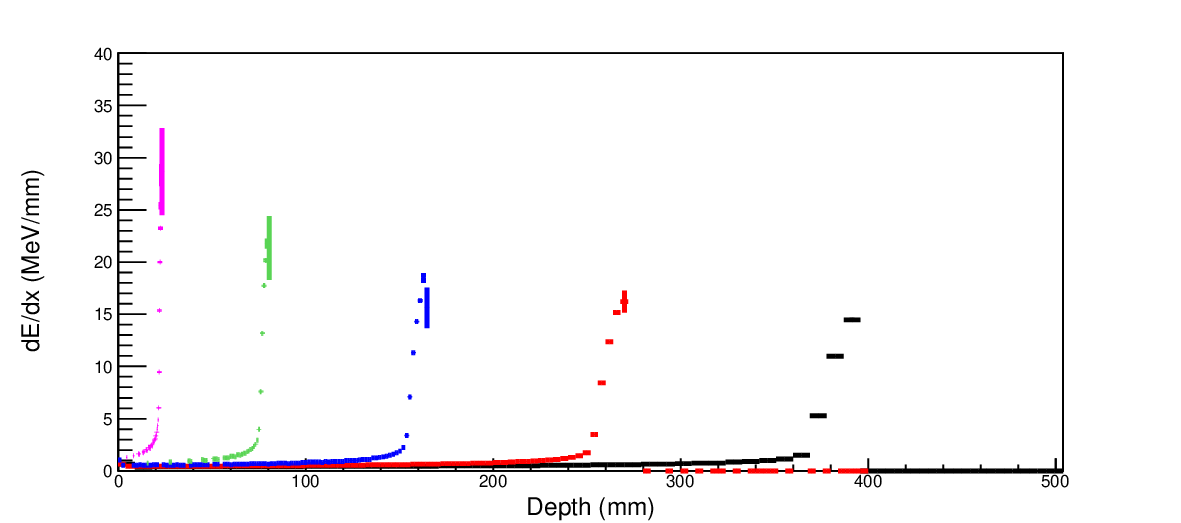}
\centering\includegraphics[height=.35\textheight]{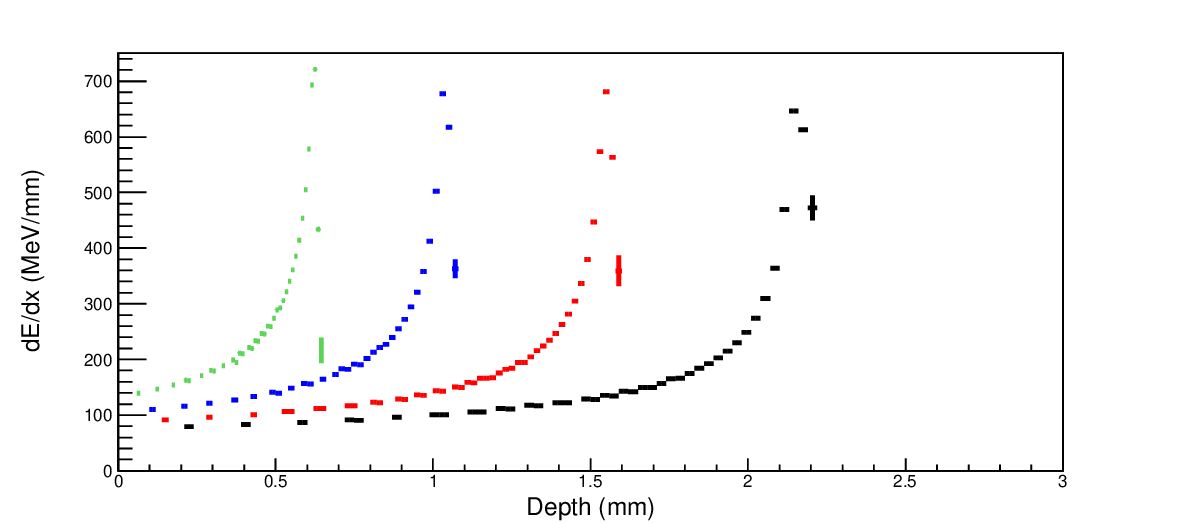}

\caption{The top graph shows Bragg peak of protons in water, where pink, green, blue, red and black curve represents protons of 50, 100, 150, 200 and 250 MeV energy respectively. The bottom graph shows Bragg peak of carbon-12 ion in water, where green, blue, red and black curve represents carbon-12 ion of 150, 200, 250, and 300 MeV energy respectively.}\label{figTwo}
\end{figure*}

\begin{figure*}
\centering\includegraphics[height=.35\textheight]{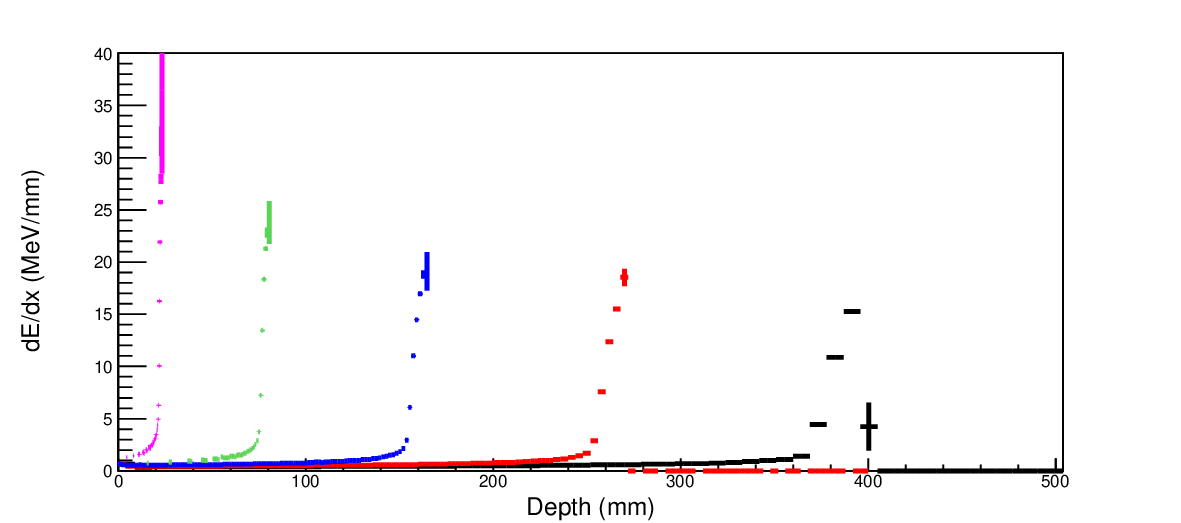}
\centering\includegraphics[height=.35\textheight]{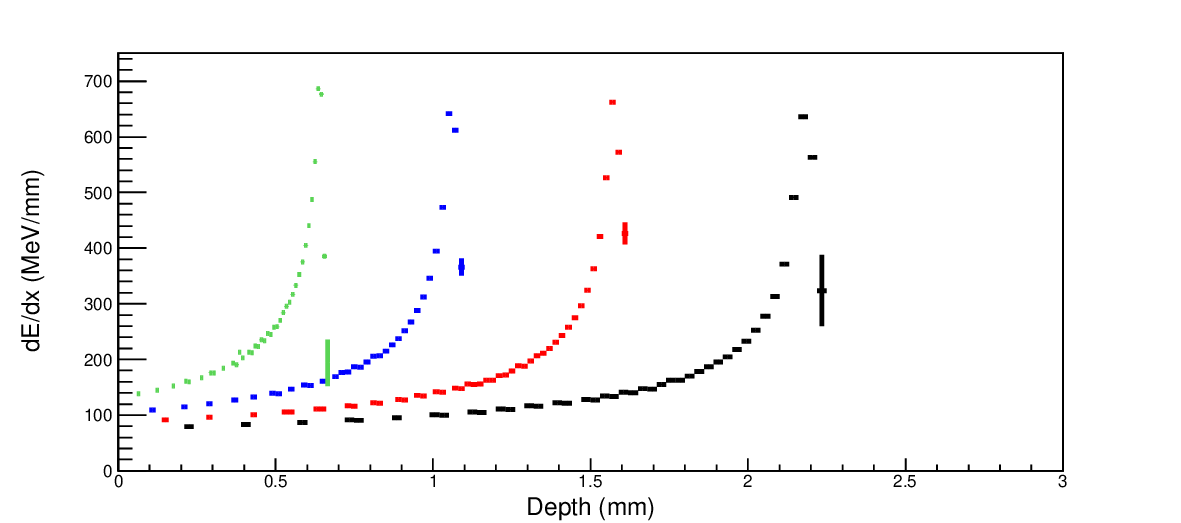}
\caption{The top graph shows Bragg peak of protons in soft tissue, where pink, green, blue, red and black curve represents protons of 50, 100, 150, 200 and 250 MeV energy respectively. The bottom graph shows Bragg peak of carbon-12 ion in soft tissues, where green, blue, red and black curve represents carbon-12 ion of 150, 200, 250, and 300 MeV energy respectively.}\label{figThree}
\end{figure*}

\begin{figure*}
\centering\includegraphics[height=.35\textheight]{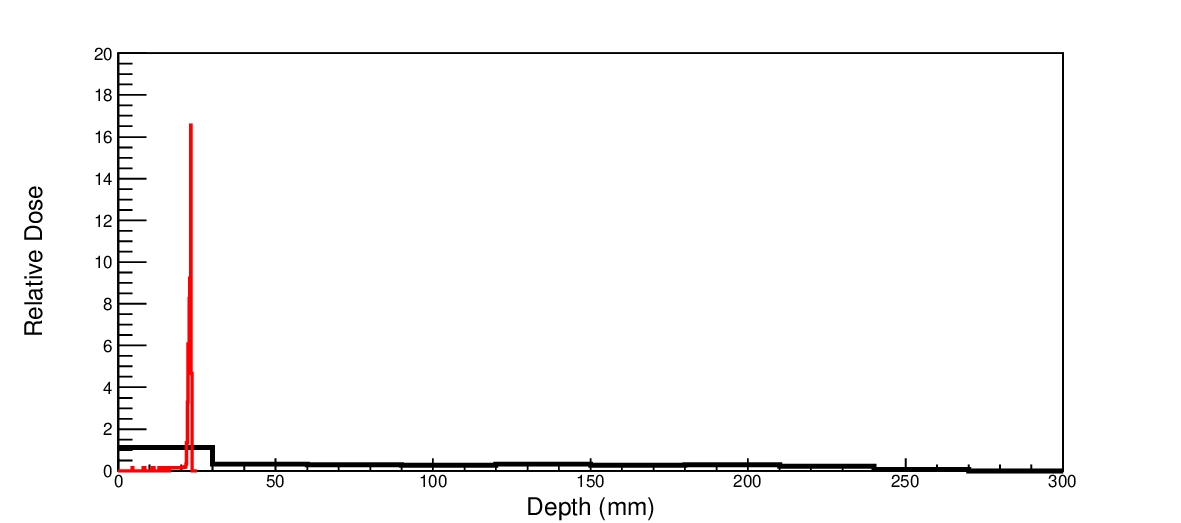}
\caption{The graph shows the relative dose of 50 MeV protons in water (red curve) and photons in water (black curve) of 50 MeV energy.}\label{figFour}
\end{figure*}

The simulation of 10,000 protons was done in water (density = 1.0 $g/cm^{3}$) and soft tissue (density = 0.9869 $g/cm^{3}$) volume. The figure~\ref{figTwo} (top graph) shows the energy deposition of protons across the depth of the water volume. These protons were shot from x = y = z = 0 cm position in a water volume. The pink, green, blue, red and black curve represents protons of 50, 100, 150, 200, 250 MeV energy respectively. In this figure the Bragg peak shows that the particle deposits energy constantly along its track but at the end of the trajectory it is maximum. This is because the energy varies inversely the square of its velocity, hence there is an increase in the interaction cross section as the protons energy decreases. Therefore just before stopping completely, protons deposits most of its energy causing Bragg's peak. A similar behaviour was observed for carbon-12 ion (bottom graph) in which green, blue, red and black curve represents carbon-12 ion shot with energy of 150, 200, 250, 300 MeV respectively. For carbon-12 ion G4EmStandardPhysics-option3, PhysListEmStandardNR and G4HadronElasticPhysics physics lists have been incorporated in our analysis. The Bragg peak is desirable for tumor/cancer treatment as after this peak, the radiation dose rapidly falls to zero. This allows for precise targeting of the tumor/cancer with less impact on surrounding healthy tissues. We obtained the similar results of protons and carbon-12 ion in soft tissue as shown in figure~\ref{figThree}. The top graph of figure shows Bragg peak of protons in soft tissue, where pink, green, blue, red and black curve represents protons of 50, 100, 150, 200, 250 MeV energy respectively. The bottom graph of the figure~\ref{figThree} shows Bragg peak of carbon-12 ion in soft tissue, where green, blue, red and black curve represents carbon-12 ion of 150, 200, 250, 300 MeV energy respectively. Also, note that the y-axis of figure~\ref{figTwo} and figure~\ref{figThree} are different. A comparison of protons with gamma photons has been shown in the figure~\ref{figFour}. The physics lists such as PhysListEmStandard, EMLivermore, EMPenelope, G4EmStandardPhysics-option3 and PhysListEmStandardNR have been used for gamma photons analysis. Therefore the Rayleigh and Compton scattering, photoelectric effect, gamma conversion and pair production processes have been incorporated in the analysis. The figure shows relative dose which the ratio of energy deposited to the energy of protons has been shown. The red curve in the graph shows the relative dose of 50 MeV protons in water and black curve shows gamma photons of 50 MeV energy in water. The comparison was done to show that protons stop after deposition of its maximum energy making a Bragg's peak hence the energy will be deposited to tumor/cancer only whereas gamma photons doesn't stop after deposition of its energy to tumor/cancer and hence effecting the healthy tissues/cells too.

\section{Summary and Conclusions}
\label{concl}

In the radiation therapy \cite{citation28}--\cite{citation34}, some of the radiation energy is absorbed by normal
organs because the radiation energy is continuously absorbed while the energy
penetrates normal tissues to reach the cancerous tissues. Therefore, a large dose of radiation is delivered to normal tissues when treating tumors/cancer deep in the body. In radiation therapy using x-rays, its basic method is to shoot radiation from several directions to spread the radiation to the surrounding normal tissues. Although only a small amount of radiation is involved, the radiation is widely distributed in the surrounding normal tissues. With proton therapy, the radiation disappears immediately after the {\it Bragg peak} \cite{citation8}. Therefore, the normal tissues surrounding the cancerous tissues are not exposed to radiation. In addition, it is possible to significantly reduce side effects during and after treatment. The simulations offers a more efficient and cost-effective alternative to traditional experimental methods for dose optimization. It streamlines the treatment planning process, potentially reducing both time and resources required for therapy.

The results shown here are crucial for understanding the interactions of protons with water and soft tissue. It was demonstrated that the Bragg peak was obtained for protons upto the minimum energy of 50 MeV whereas with carbon-12 it was obtained with minimum of 150 MeV energy. Also it was observed that for 150 MeV of protons and carbon-12 in water the range was 15.2 cm and 0.065 cm respectively. Due to which protons show much better range and hence energy deposition than carbon-12. These studies can lay foundation in treatment of tumor/cancer as convectional methods like surgeries/chemotherapy does much damage to the patients.

In the end we would say that a rigorous computational analysis has been done to achieve significant outcomes. The proton simulations demonstrated a remarkable improvement in the dose calculations compared to carbon-12 ion studies. These results are critical for ensuring optimal treatment efficacy while minimizing adverse effects on healthy tissues. Tailoring the radiation dosage to individual patient characteristics can lead to improved outcomes and reduced toxicity \cite{citation35}--\cite{citation40}. As a next-generation therapy i.e., proton therapy are expected to treat more cancer patients in the future.

\section*{Acknowledgement}
The author would like to thank the center of excellence in computational physics, department of physics, Chandigarh University for help and support.

\end{document}